\documentstyle[aps,version2,prl]{revtex}

\begin{document}
\begin{title}
Damping mechanism for the strongly renormalized $c$-axis charge transport

in high-$T_c$ cuprate superconductors
\end{title}
\author{Jae H. Kim, H. S. Somal, and D. van der Marel}
\begin{instit}
Laboratory of Solid State Physics, University of Groningen,

Nijenborgh 4, 9747 AG Groningen, The Netherlands
\end{instit}
\moreauthors{A. M. Gerrits, and A. Wittlin}
\begin{instit}
High Field Magnet Laboratory, University of Nijmegen,

Toernooiveld 1, 6525 ED Nijmegen, The Netherlands
\end{instit}
\moreauthors{V. H. M. Duijn, N. T. Hien, and A. A. Menovsky}
\begin{instit}
Van der Waals-Zeeman Laboratory, University of Amsterdam,

Valckenierstraat 65, 1018 XE Amsterdam, The Netherlands
\end{instit}
\receipt{}
\begin{abstract}
We analyze the $c$-axis infrared reflectivity
of La$_{1.85}$Sr$_{0.15}$CuO$_4$ single crystals.
The plasma edge near 6 meV,
observed below $T_c$,
is due to Cooper-pair tunneling.
This low value of the plasma edge is consistent with
the $c$-axis plasma frequency ($\nu_p$)
obtained from LDA calculations ($>0.1$ eV)
if we take into account that
the single-particle charge transport along the $c$ axis is strongly
incoherent both above and below $T_c$.
We find no evidence
for a reduction of the $c$-axis scattering rate ($\gamma$) below $T_c$.
Our investigation suggests
$h\gamma>h\nu_{p}\gg 3.5k_BT_c$,
which is exactly opposite to the clean limit.
\end{abstract}
\vspace{1\baselineskip}\\
Materials Science Center Internal Report Number VSGD.94.6.1
\vspace{1\baselineskip}\\
\twocolumn
One of the peculiar aspects of the high-$T_c$ cuprate superconductors is
the large anisotropy related to their layered structure.
Although the anisotropy following from a conventional Fermi-liquid approach
based on LDA (local density approximation) band structure calculations,
is of the order 10 in most of these compounds,
experimentally it is found to be much larger,
as follows from {\em e.g.} the London penetration
depth and the normal-state resistivity.  In the clean limit the London
penetration depth $\lambda_L$ is given by
$c^2/\lambda_L^2=4 \pi n e^2/m^*$
and usually the large anisotropy of $\lambda_L$ is attributed to
a large mass enhancement \cite{Tamasaku,Shibauchi}.
A mass enhancement could follow from {\em e.g.} spin-charge separation
\cite{pwaspincharge,tmrice,mjrice}.
Alternatively, elastic \cite{Graf} or inelastic \cite{Rojo}
Fermi-liquid corrections, as well as spin-charge separation,
may cause a strongly incoherent charge transport along the $c$ axis.
No plasma pole is found in the normal state, whereas in the
superconducting state in some of the materials a zero-crossing
of the real part of the dielectric function ($\epsilon^{\prime}$) is found
at an energy two orders of magnitude below the value expected from
LDA band structure calculations. In this
paper we shall argue that the discrepancy results not from a strong
mass-enhancement, but from a strong  quasi-particle damping for
$\vec{E} // c$, both in the normal and in the superconducting state.

Single crystals of La$_{2-x}$Sr$_{x}$CuO$_{4-\delta}$
with nominal Sr contents of 10 and
15 \% were grown by the traveling-solvent floating-zone method \cite{Duijn}.
Electron probe micro analysis showed the correct phase with
no contamination or inclusion of second phases, and a
homogeneous distribution of strontium. The crystals have sharp superconducting
transitions (35 K at optimal doping, with 90 \% of the transition within
3 K as determined from AC susceptibility measurements at 3 Gauss, 90 Hz).

Neutron diffraction measurements demonstrated that
the as-grown boules contain large single crystalline grains.
Samples for reflectivity measurements were
cut perpendicular to the $b$ axis
and were mechanically polished to optical quality.
The surface in the $ac$ plane was approximately 4$\times$7 mm$^2$.
X-ray Laue backscattering showed a
correct orientation with sharp, well-defined single
diffraction spots.

Polarized infrared reflectivity spectra were measured
in two different laboratories between
10 and 10000 cm$^{-1}$ at temperatures between 10 and 300 K using
rapid-scan Fourier-transform spectrometers equipped with a
continuous-flow cryostat. The data obtained in the two laboratories
on nominally identical crystals are in a good agreement with each other,
and with those of Tamasaku {\em et al.} \cite{Tamasaku}.

In the inset of Fig. 1 we display the reflectivity
of a La$_{1.85}$Sr$_{0.15}$CuO$_4$ single crystal
with $\vec{E} //\vec{c}$ in the normal state at 33 K. The spectrum
is typical for an ionic insulator. At low frequencies there
is a Hagen-Rubens type of upturn, $R(\nu)=1-2(\sigma_0/\nu)^{1/2}$,
characteristic of a poor conductor,
where Re$\sigma(\nu)$ is dominated by
$\sigma_0(1+\nu^2/\gamma^2)^{-1}$,
with $\sigma_0 \approx 7\;\Omega^{-1}\mbox{cm}^{-1}$.
A closer inspection, using
a Drude-Lorentz model with a Drude oscillator
and 4 phonons (see Fig. 1 caption),
indicates a featureless and almost constant value of the electronic
part of $\sigma(\nu)$: A good fit in the
range of 20 to 650 cm$^{-1}$ can only be obtained (see Fig. 1 inset)
if $\gamma$ is taken to be larger than $0.5$ eV. With these parameters
a zero-crossing of $\epsilon^{\prime} =
\tilde{\epsilon}_{\infty}-2\sigma_0\gamma(\nu^2+\gamma^2)^{-1}$
occurs for $\nu^2 <0$, and hence
the plasmon has a diffusive pole in the normal state.
By integrating the corresponding expression for Re$\sigma(\nu)$, we can put a
lower
bound on $\nu_p$ of $\sqrt{2\sigma_0\gamma}=0.15$ eV. Interestingly
this is much closer to estimates of the $c$-axis plasma frequency based
on the single-electron effective mass in this solid. From such
an RPA-LDA analysis for the La$_2$CuO$_4$ system one obtains a screened
$ab$-plane plasma frequency of 2 eV, well above the onset of the
interband electron-hole continuum at 0.5 eV, and a screened $c$-axis
plasma frequency at 0.35 eV, {\em i.e.},
below the interband electron-hole continuum \cite{czyzyk}.
It is interesting to notice that
experimentally the situation {\em vis \'a vis}
Landau damping is reversed: In the $ab$-plane dielectric function
a clear, though strongly damped, plasma pole is observed near 1 eV
\cite{Kim.1991}, whereas
a plasmon is absent in the $c$-axis dielectric function due
to overdamping. Such loss of coherence is also born out by
confinement of holons to the planes\cite{confinement},
due to the fact that an interlayer hopping process requires
a spinon-holon combination as an intermediate
state.

Let $E_g^*$ be the {\em effective} value of the gap, or gap-distribution.
We will assume $E_g^*=3.5 k_BT_c$, which becomes exact in the limit of a
weak coupling $s$-wave superconductor.
In the superconducting state we should make a distinction between
the high- and low-frequency parts of the spectrum,
far above and below $E_g^*$
($\approx 80\mbox{ cm}^{-1}$ for La$_{1.85}$Sr$_{0.15}$CuO$_4$).
The former remains relatively unaffected by the transition,
whereas for the latter range of frequencies
we expect to find
temperature dependence in both the real and imaginary parts of
the electronic contribution to the conductivity.
For example, for an isotropic
$s$-wave BCS superconductor one expects a decrease of $\sigma(\nu)$ for
$\nu \approx E_g^*/2$ as indicated in Fig. 2. In addition there may
be a strong decrease of $\gamma$ for $T<T_c$ as has been
reported for the in-plane infrared \cite{Romero,Nuss} and microwave
\cite{Bonn.1993} response of high-$T_c$ superconductors, and
the $c$-axis infrared response of La$_{2-x}$Sr$_{x}$CuO$_4$\cite{Tamasaku}.

We analyze our data in the superconducting state using the two-fluid
Gorter-Casimir model at frequencies up to 100 cm$^{-1}$, {\em i.e.},
with the model dielectric function\cite{twofluid}
\begin{displaymath}
\epsilon=\tilde{\epsilon}_{\infty}-\frac{\nu_{\phi}^2}{\nu^2}
+i\frac{2\sigma_0}{\nu(1-i\nu/\gamma)}.
\end{displaymath}
The dashed curve in Fig. 1 corresponds
to the fit-parameters used by Tamasaku {\em et al.} for 8 K \cite{Tamasaku}.
Note that with their set of parameters in the superconducting
state nearly all oscillator strength under $\sigma$ collapses into
the $\delta$-function at the origin. This essentially corresponds
to the clean limit.
A much better fit is obtained if the spectra are modeled with
a broad and featureless electronic component of $\sigma$, similar to the normal
state.
Initially we varied all parameters. At all temperatures
the best fits indicate that $\nu/\gamma \ll 1$, and only small and unsystematic
changes of $\tilde{\epsilon}_{\infty}$ with temperature are obtained.
Therefore we repeated the analysis with $\gamma \rightarrow \infty$,
and replacing $\tilde{\epsilon}_{\infty}$ by the
sum of $\epsilon_{\infty}$ and the oscillator strengths of Fig. 1. To
estimate the influence of dispersion of the optical phonons, we
also did this analysis using the full set of phonon parameters of Fig. 1.
The temperature dependence
of the remaining two parameters ($(\nu_{\phi}(T)/\nu_{\phi}(0))^2$
and $\sigma_0(T)/\sigma_0(T_c)$, with $\nu_{\phi}(0)=180\mbox{ cm}^{-1}$ and
$\sigma_0(31\mbox{ K})=6.7\;\Omega^{-1}\mbox{cm}^{-1}$ are displayed in Fig. 2,
and the reflectivity data together with the fitted curves
are shown for three different temperatures in Fig. 1.

Also shown in Fig. 2 are the results of model calculations for the
dirty limit of isotropic $s$-wave \cite{Mattis} and $d$-wave
\cite{Won} superconductors, assuming standard ratios of $E_g^*/T_c$ for these
two models, and assuming a temperature-independent scattering rate. The
conductivity was calculated for $\nu=50\mbox{ cm}^{-1}$. This frequency
is close both to BCS ratio $ 3.5k_BT_c = 80\;\mbox{cm}^{-1} $
and to the gap value deduced from neutron scattering \cite{mason}.
As a result we observe a steep drop
of $\sigma$ for $T<T_c$, and the change of analytical behavior
at the temperature where $E_g(T)=h\nu$. For $d$-wave pairing the
calculated temperature dependence of $\sigma$ is smooth, and
quite close to the experimentally determined values. If we would
assume that in addition there
is a drop of the decay rate when the gap opens, the theoretical
curves would have an even steeper temperature dependence below $T_c$.
For this reason
we attribute the decrease of $\sigma$ for $T<T_c$ exclusively to the
opening of a gap, {\em i.e., without} assuming a change of the
electronic decay rate. For $\sigma_0(T)$ a reasonable agreement exists
with the $d$-wave model, whereas there is no good agreement with the
$s$-wave model. For $f_s(T)=(\nu_{\phi}(T)/\nu_{\phi}(0))^2$ the
situation is reversed: $f_s$
follows roughly a $1-(T/T_c)^4$ law, which differs strongly from
the $d$-wave pairing model, and lies somewhat above the $s$-wave pairing
curve. For $\sigma_0(T)$ the comparison with the
$s$-wave model is slightly improved
if we postulate the existence of constant
background (possibly spurious) conductivity in our samples.
Interestingly,
the jump in $\chi^c(T)/\chi^{ab}(T)$,
as recently reported by Shibauchi {\em et al.}\cite{Shibauchi},
is very well reproduced within this picture,
as well as the difference between $\lambda_{ab}$ and $\lambda_c$,
if we assume a clean limit picture for the $ab$-plane response.

 From a direct inspection of $\sigma(\nu)$ obtained from
Kramers-Kronig analysis (Fig. 1 lower panel) we find no signature of
a gap in the data\cite{Gerrits}, possibly due to limitations in signal-to-noise
ratio
and the presence of a tail of the optical phonon at $240\;\mbox{cm}^{-1}$.
As is clear from the calculated conductivity (the dashed curve)
based on Mattis-Bardeen theory,
even in the case of an isotropic $s$-wave superconductor
a gap would be hard to distinguish. In the present case the difficulty arises
from the low value of $\sigma$ due to a strong damping. We notice that
the situation is reversed compared to the planar response.
In the latter case $E_g^*$ is situated in the Lorentzian tail of a
narrow Drude peak, which, due to the low value of $\sigma$, makes a gap
difficult to observe\cite{kamaras}.

In Fig. 3 we classify various regimes that can occur
depending on the relevant energy scales: First there is the pair-breaking
energy scale $E_g^*$. Second is the screened plasma frequency
$\tilde{\nu}_p=\tilde{\epsilon}_{\infty}^{-1/2}\nu_p$,
(the screening may include optical phonons depending
on how far down the plasmon is pushed due to
screening and damping effects).
Similar parameters were also used by Uemura {\em et al.} \cite{Uemura}
to classify clean-limit superconductors corresponding
to their Fermi energy ($\propto h\nu_p$ for a 2D Fermi liquid). We extend
this classification with the
the scattering rate $\gamma$ (using $2\gamma\sigma_0=\nu_p^2$).
We scale both $E_g^*$ and $\gamma$ to ${\tilde \nu_p}$.

The region labeled `CL' corresponds to the clean limit where
$\gamma \ll E_g^*$, and this region contains
the `exotic' superconductors considered by Uemura {\em et al.} \cite{Uemura},
in particular
the high-$T_c$ cuprate superconductors characterized by their
in-plane electrodynamical properties.
Most classical superconductors fall in the region labeled `MB',
where the infrared properties can be described in the context of
dirty-limit Mattis-Bardeen theory. In the CL and MB regions
a plasmon is found well above $E_g^*$ in
the normal as well as in the superconducting state.

In the regions labeled `OP' and `RP',
the plasmon is overdamped in the normal state,
as with $\gamma > \tilde{\nu}_p$ a zero crossing
of $\epsilon^{\prime}$ no longer exists.
In the superconducting state the `missing area' of the conductivity
due to the opening of the gap
is transferred to the $\delta$-function at $\nu=0$,
so that $\epsilon^{\prime}$ crosses zero at
$\tilde{\nu}_p\sqrt{\pi E_g^*/2h\gamma}$.
Unless this frequency is below the gap,
the plasma mode will be damped (OP region). Hence the requirement for the
existence of a long-lived collective mode is, that
$E_g^*/\tilde{\nu}_p > \pi \tilde{\nu}_p/2\gamma$. This `phason' mode has been
described by Doniach and Inui \cite{Doniach} assuming Josephson
coupling between
the layers. Using Anderson's expressions\cite{AndersonHiggs}
for collective modes in
the superconducting state we have recently shown\cite{marel}, that the plasmon
gradually changes from a collective oscillation
of quasi-particles for $\nu\gg E_g^*$,
to an oscillating current carried by the condensate for $\nu\ll E_g^*$.
This region is labeled `RP' (re-entrant phason).
Our investigation indicates that the $c$-axis response of
La$_{1.85}$Sr$_{0.15}$CuO$_4$ lies in the RP region.
Although the $c$-axis response data for other high-$T_c$ materials
are very limited, the far infrared results of Noh {\em et al.} \cite{noh}
and of Tajima {\em et al.} \cite{tajima}
indicate that also the {\em c}-axis response
of Nd$_{1.85}$Ce$_{0.15}$CuO$_{4-y}$ and of Bi$_2$Sr$_2$CaCu$_2$O$_{8+x}$
should be placed in the RP region. In both these materials the
$c$-axis plasmon, attributed to a sphere resonance by
Noh {\em et al.} \cite{noh}, is found at very low energy
of approximately 10 cm$^{-1}$,
and in Bi$_2$Sr$_2$CaCu$_2$O$_{8+x}$ a very strong damping of
$c$-axis electronic response \cite{tajima} is observed.

The two regions labeled CLP (`clean local pair')
and OLP (`overdamped local pair')
correspond to clean and dirty
limit superconductors with a pair-breaking energy larger than the
plasma frequency. In such systems the phase transition
is induced by thermal activation of phasons, while the pairs
remain intact up to a higher de-pairing temperature scale.
This corresponds to Bose-Einstein condensation of local pairs.
According to the analysis of Tamasaku {\em et al.} \cite{Tamasaku}
the $c$-axis phason of La$_{2-x}$Sr$_{x}$CuO$_4$ would be located in the
CLP region.

We conclude that the absence of a $c$-axis plasmon
in the normal state of high-$T_c$ cuprate superconductors
results from an anomalously strong damping
of the transport perpendicular to the CuO$_2$ planes. This assignment
also alleviates the discrepancy by two orders of magnitude existing in
the literature between the plasma frequency obtained from
optical experiments and that from the single-electron interlayer-hopping
matrix element. The appearance of a $c$-axis phason mode, as
observed in the superconducting state, is a direct consequence of
the anomalously strong damping of the charge-carrier transport perpendicular
to the planes.
The decrease in $\sigma$ for $T<T_c$ can be fully accounted for by
the opening of a gap, the presence of which is a necessary consequence
of the $f$-sum rule. There is no indication for a sudden decrease
of the optical decay rate for $\vec{E}//\vec{c}$ when $T<T_c$.

{\em Acknowledgements}
We gratefully acknowledge stimulating discussions with M. J. Rice, and
useful comments by T. M. Rice and P. W. Anderson during the preparation
of this manuscript.
This investigation was supported by the Netherlands Foundation for
Fundamental Research on Matter (FOM) with financial aid from
the Nederlandse Organisatie voor Wetenschappelijk Onderzoek (NWO).

\figure{
Upper panel: Reflectivity of La$_{1.85}$Sr$_{0.15}$CuO$_4$
at 10 (squares), 20 (circles), and 27 K (triangles)
with least-square fitted curves.
Dashed curve: fit-parameters for 8 K taken from
Tamasaku {\em et al.} \cite{Tamasaku}.
Inset: Reflectivity at 33 K (triangles) with
Drude-Lorentz fit. The parameters are
$\epsilon_{\infty}=4.75$,
$\sigma_0=7\;\Omega^{-1}\mbox{cm}^{-1}$,
$\gamma >4600\;\mbox{cm}^{-1}$,
so that $\nu_p>1200\;\mbox{cm}^{-1}$.
The TO phonons were parametrized with
\{$\omega_{i}$ (cm$^{-1}$),
$\gamma_{i}$ (cm$^{-1}$),
S$_{i}$\} ($i=1\cdots4$):
\{241.5, 24.8, 17.6\},
\{312.0, 3.0, 0.019\},
\{351.0, 3.0, 0.044\},
\{494.1, 12.8, 0.33\}.
Lower panel: Optical conductivity obtained from Kramers-Kronig
analysis at 10, 20, and 27 K (solid symbols, from bottom to top).
Chained curve: -Im$\epsilon^{-1} \times 13.0 \Omega^{-1}\mbox{cm}^{-1}$.
Open symbols: conductivities with the chained curves subtracted.
Dashed curve: calculation based on the Mattis-Bardeen
model, taking into account the phonon contribution.
\label{fig1}}
\figure{
Temperature dependence of $\sigma_0$ (solid symbols)
and $\nu_\phi^2$ (open symbols) corresponding to the fits in Fig. 1.
Squares: sample 1. Circles and triangles : sample 2.
Open and closed squares and open circles: with a constant
${\tilde \epsilon}_{\infty}$. Open and closed triangles:
with full frequency dependence of phonon oscillators
in ${\tilde \epsilon}_{\infty}$.
Solid circles: conductivities at 50 cm$^{-1}$, after
subtracting the loss function peaks as indicated in the lower panel
of Fig. 1. Dashed curve: $1-(T/T_c)^4$.
Chained and solid curves: simulation for $s$- and $d$-wave pairing,
respectively. Dotted
curve: $s$-wave pairing plus constant background conductivity.
The simulations of $\sigma_s/\sigma_n$ are at $\nu = 50\;\mbox{cm}^{-1}$.
\label{fig2}}
\figure{
Phase diagram in the $(\gamma/{\tilde \nu_p})$-$(E_g^*/h{\tilde \nu_p})$ plane
\cite{Kim.1994}.
\label{fig3}}

\end{document}